\documentstyle[12pt]{article}
\begin{document}
\newcommand{\siml}{\stackrel{<}{\sim}}
\newcommand{\simg}{\stackrel{>}{\sim}}
\newcommand{\lleq}{\stackrel{<}{=}}
\baselineskip=1.333\baselineskip

\noindent
\begin{center}
{\large\bf 
Nonextensive aspects of small-world networks
} 
\end{center}

\begin{center}
Hideo Hasegawa
\footnote{e-mail:  hasegawa@u-gakugei.ac.jp}
\end{center}

\begin{center}
{\it Department of Physics, Tokyo Gakugei University  \\
Koganei, Tokyo 184-8501, Japan}
\end{center}
\begin{center}
\today
\end{center}
\thispagestyle{myheadings}

\begin{center} 
{\bf Abstract}   \par
\end{center}
Nonextensive aspects of the degree distribution in
Watts-Strogatz (WS) small-world networks, $P_{SW}(k)$,
have been discussed in terms of a generalized Gaussian 
(referred to as {\it $Q$-Gaussian})
which is derived by the three approaches:
the maximum-entropy method (MEM),
stochastic differential equation (SDE), and
hidden-variable distribution (HVD).
In MEM, the degree distribution $P_Q(k)$ 
in complex networks has been obtained from $Q$-Gaussian
by maximizing the nonextensive information entropy
with constraints on averages of $k$ and $k^2$
in addition to the normalization condition. 
In SDE, $Q$-Gaussian is derived from Langevin equations
subject to additive and multiplicative noises.
In HVD, $Q$-Gaussian is
made by a superposition of Gaussians for random networks
with fluctuating variances, in analogy to superstatistics.
Interestingly, 
{\it a single} $P_{Q}(k)$ may describe, with an accuracy of
$\mid P_{SW}(k)-P_Q(k)\mid \siml 10^{-2} $,
main parts of degree distributions of SW networks, 
within which about 96-99 percents of all $k$ states
are included.
It has been demonstrated that
the overall behavior of $P_{SW}(k)$ including its tails
may be well accounted for
if the $k$-dependence is incorporated into the entropic index in MEM, 
which is realized in microscopic
Langevin equations with generalized multiplicative noises.

\vspace{0.5cm}


\vspace{1.0cm}
\noindent
{\it Corresponding Author}

\noindent  
Hideo Hasegawa \\ 
Department of Physics, Tokyo Gakugei University \\ 
4-1-1 Nukui-kita machi, Koganei, Tokyo 184-8501, Japan \\
Phone: 042-329-7482, Fax: 042-329-7491 \\
e-mail: hasegawa@u-gakugei.ac.jp



\newpage

\section{INTRODUCTION}


In the last several years, 
there is an increased interest in
two subjects of statistical physics:
(1) {\it nonextensive statistical mechanics} 
\cite{Tsallis88}-\cite{NES}
and (2) {\it complex networks} 
\cite{Wat98}-\cite{Dor02}.
As for the first subject (1),
Tsallis has developed the nonextensive
statistics (NES), proposing the generalized entropy 
given by \cite{Tsallis88}\cite{Tsallis98}
\begin{equation}
S_q = \frac{\sum_i p_i^q-1}{1-q},
\end{equation}
where $q$ denotes the entropic index
and $p_i$ the probability distribution
of the state $i$.
It is noted that in the limit of $q=1$,
$S_q$ reduces to the entropy in the Boltzman-Gibbs
statistics given by
\begin{equation}
S_1 = - \sum_i p_i \:{\rm ln}\: p_i.
\end{equation}
The NES has been successfully
applied to a wide range of nonextensive systems
including physics, chemistry, mathematics, astronomy, 
geophysics, biology, medicine, economics, 
engineering, linguistics, and others \cite{NES}.

The subject (2) has been initiated by
the seminal paper of Watts and Strogatz \cite{Wat98}.
Since then considerable study has been made
on complex networks \cite{Alb02}\cite{Dor02}. 
Existing and proposed
networks are classified into two types:

\noindent
(i) Networks where the number of
nodes is constant and where the 
degree distribution $P(k)$ for
a given node to be connected to $k$ neighbors, has
a maximum at or near $<k>$ with finite width
where $<\cdot>$ denotes the average over $P(k)$. 
This type includes random \cite{Erdos60}
and small-world networks
\cite{Wat98}\cite{Str01}.

\noindent
(ii) Networks where the number of nodes is not stationary
and where the distribution varies between
the exponential and power forms.
This type includes scale-free (SF) networks,
which were originally proposed 
by Barab\'{a}si and Albert \cite{Bara99}
with a growth of nodes
and their preferential attachments.
Many models and mechanisms have been proposed 
not only for growing but also for non-growing SF networks 
with geographical and non-geographical structures.

It is possible that the two subjects (1)
and (2) are intimately related
\cite{Wilk04}-\cite{Abe05}.
By using the nonextensive information entropy
as given by Eq. (1),  
Wilk and Wlodarczyk have 
shown that degree distribution 
$P(k)$ of SF networks which belongs to type (ii), 
is given by \cite{Wilk04}\cite{Wilk05}
\begin{equation}
P_{SF}(k) \propto \left[1-(1-q)\left(\frac{k}{\mu}\right)\right]
^{\frac{q}{1-q}},
\end{equation}
where $\mu=<k>$ and $<\cdot>$ denotes the average over $P(k)$. 
The expression (3) is universal because
in the limit of $q=1$ it reduces to the exponential
distribution given by $P(k) \propto {\rm e}^{-k/\mu}$,
and for $k \gg \mu/(q-1)$
it yields the power law given by 
$P(k) \propto k^{-q/(q-1)}$.
Thus the expression (3) successfully 
accounts for degree distribution in SF networks.
Soares {\it et al.} \cite{Soares04} have
proposed the two-dimensional growth model
for SF networks where 
the degree distribution is given by Eq. (3)
with $\mu$ and $q$ expressed in terms of model
parameters.
Thurner and Tsallis \cite{Thurner05}
have discussed scale-free gas-like networks,
showing their nonextensive aspects. 
It has been pointed out by
Abe and Thurner \cite{Abe05}
that degree distribution in scale-free networks 
may be generated by superimposing Poisson distributions 
relevant to random networks \cite{Erdos60} with 
a proper hidden-variable distribution, 
just as in superstatistics \cite{Beck03}.

As for the type (i) networks,
degree distribution $P(k)$ of random network is given by
the binomial distribution \cite{Erdos60}:
\begin{eqnarray}
P_{B}(k)&=&C^{N-1}_{k} c^k\:(1-c)^{N-1-k}, 
\end{eqnarray}
where $c=\mu/N$ and $C_n^m=m!/n!\:(m-n)!$.
In the limit of large $N$, Eq. (4) becomes the Poisson
distribution given by
\begin{eqnarray}
P_{P}(k)=P_P(k;\mu)&=&\frac{\mu^k}{k!} \;e^{-\mu}.
\end{eqnarray}
On the contrary,
$P(k) $ of Watts-Strogatz (WS) 
small-world (SW) networks \cite{Wat98}
is given by \cite{Barrat00}
\begin{eqnarray}
P_{SW}(k)&=& \sum_{n=0}^{f(k,\mu)}
C^{\mu/2}_{n}\:p^{\mu/2-n}(1-p)^n
P_P(k-\mu/2-n; p\mu/2), \nonumber \\
&& \hspace{5cm}\mbox{(for $k \geq \mu/2$ and large $N$) }
\end{eqnarray}
where $f(k,\mu)={\rm min}(k-\mu/2,\mu/2)$
and $p$ ($=2 N_r /N \mu$) denotes a randomness parameter
for random rewirings of $N_r$ edges from a regular network
\cite{Wat98}.
Figure 1 shows $P_{SW}(k)$
with $N=100$ and $\mu=10$ for various randomness
$p$ \cite{Hasegawa04}. 
Results shown in Fig. 1
are obtained by simulations with 1000 trials
after Ref.\cite{Wat98},
because the analytic expression given by Eq. (6) is valid
only for $k \geq \mu/2$ and large $N$.
For $p=0$ (regular networks), $P_{SW}(k)$ becomes
a delta-function at $k=\mu$.
With more increasing $p$, $P_{SW}(k)$ 
has a peak at $k=\mu$ with a wider width.
As a comparison, we show by open circles in Fig. 1, 
$P_B(k)$ of random networks given by Eq. (4).
Note that even for $p=1$, $P_{SW}(k)$ of 
the WS-SW network
is different from $P_B(k)$ of random networks because
the former is not locally equivalent to the latter
with some memory of the starting regular network  
\cite{Barrat00}.

We suppose that $P_{SW}(k)$ shown in Fig. 1
may be described or approximated 
by a simple function like Eq. (3)
if the nonextensive property of SW networks
is properly taken into
account. The purpose of present paper is to
investigate such a possibility. This is indeed possible
in an approximate sense,
as will be demonstrated in the present paper.

The paper is organized as follows. 
By using the maximum-entropy method (MEM)
with constraints
on $<k>=\mu$ and $<k^2>=\rho$ besides the normalization condition,
we get in Sec. 2, the generalized
Gaussian with a maximum at $k=\mu$
(hereafter referred to as {\it $Q$-Gaussian}). 
In contrast, a conventional
$q$-Gaussian distribution which has been widely 
employed in many subjects such as self-gravitating 
stellar systems \cite{Taruya03} and ideal gas \cite{Lima05},
has a peak at the zero value. 
In Sec. 3, we discuss the stochastic-differential equation (SDE)
by using Langevin equation with additive and 
multiplicative noises.
In Sec. 4, SW networks are 
constructed by hidden-variable distribution (HVD).
An analysis of degree distributions of SW 
networks is made in terms of
$Q$-Gaussian in Sec. 5.  
Section 6 is devoted to more detailed analysis, including
the $k$-dependence in the entropic index in MEM,
which is shown to be realized in Langevin equations
with generalized multiplicative noises.
Our conclusions are presented in Sec. 7.

Before going to Sec. 2, we summarize
various distributions used in this paper as follows.
$P_{SF}$(k), SF networks;
$P_{SW}$(k), SW networks;
$P_B(k)$, binomial; 
$P_P(k)$, Poisson;
$P_G(k)$, Gaussian; 
$p_k^Q$, $Q$-Gaussian;
$P_{Q}(k)$, an escort probability of $p_k^Q$.

\section{Maximum-entropy method}

\subsection{Shannon information entropy}

We have adopted a network with $N$ nodes whose adjacent 
matrix is given by
$c_{ij}=c_{ji}=1$ for a coupled pair $(i, j)$
and zero otherwise.
The degree distribution $P(k)$ for a given node to have
$k$ neighbors is defined by
\begin{equation}
P(k)=\frac{1}{N} \sum_i <\delta(k-\sum_j c_{ij})>_G,
\end{equation} 
where $<>_G$ denotes the average over graphs.
The averaged coordination number $\mu$ is given by
\begin{eqnarray}
\mu&=&\frac{1}{N} \sum_{i} \sum_j <c_{ij}>_G, \\
&=& \sum_k P(k) \:k.
\end{eqnarray} 
The {\it coupling connectivity} $R$ is defined by \cite{Hasegawa04}
\begin{eqnarray}
R &=& \frac{1}{N \mu^2} \sum_i \sum_j \sum_{\ell} 
<c_{ij}c_{i\ell}>_G, \\
&=& \frac{1}{\mu^2} \sum_k P(k) \:k^2.
\end{eqnarray} 
The coupling connectivity $R$
expresses a factor for a cluster 
where the two sites $j$ and $\ell$
are coupled to the third site $i$, but the sites $j$ and $\ell$ 
are not necessarily coupled.
In contrast, the clustering coefficient $C$ defined by
\begin{eqnarray}
C&=& \frac{1}{N \mu^2} \sum_i \sum_j \sum_{\ell} 
< \:c_{ij}c_{j\ell}c_{i\ell} >_G, 
\end{eqnarray}
expresses a factor forming a cluster 
where the three sites $i$, $j$ and $\ell$
are mutually coupled \cite{Note2}.
Unfortunately $C$ cannot be expressed in terms
of $P(k)$.

First we consider the Shannon entropy given by
\begin{equation}
S= - \sum_{i} P(k)\: {\rm ln} \:P(k).
\end{equation} 
Assuming that $\mu$ and $R$ of the adopted networks
are given, we obtain $P(k)$ by MEM
with the three constraints given by
\begin{eqnarray}
\sum_k P(k)&=&1, \\
\sum_k P(k) \:k &=& \mu, \\
\sum_k P(k) \:k^2 &=& \mu^2  R \equiv \rho^2. 
\end{eqnarray}
We get
\begin{equation}
P(k) \propto e^{-(\beta k+\gamma k^2)},
\end{equation} 
where $\beta$ and $\gamma$ are Lagrange multipliers
relevant to the constraints given by Eqs. (15) and (16),
respectively. 

\subsection{Nonextensive information entropy}

Next we extend our discussion by employing
the generalized entropy first
proposed by Tsallis \cite{Tsallis88,Tsallis98}:
\begin{equation}
S_q=\frac{\sum_k p_k^q - 1}{1-q},
\end{equation} 
which reduces to the Shannon entropy 
in the limit of $q =1$.
Here $p_k$ denotes the probability distribution
for couplings of
$k$ neighbors whose explicit form will be given shortly
[Eq. (24)].
The three constraints corresponding to Eqs. (14)-(16)
are given by
\begin{eqnarray}
\sum_k p_k&=&1, \\
\sum_k P(k) \:k &=& \mu, \\
\sum_k P(k) \:k^2 &=& \mu^2  R = \rho^2, 
\end{eqnarray}
with
\begin{eqnarray}
P(k)&=& \frac{p_k^q}{c_q}, \\
c_q&=& \sum_k p_k^q,
\end{eqnarray}
where $P(k)$ is the escort probability 
expressing the degree distribution \cite{Tsallis98}.
By maximizing the entropy 
with the three constraints, we get
\begin{equation}
p_k \propto  
\left[ 1- \left(\frac{1-q}{c_q}\right) [\beta(k-\mu)
+\gamma (k^2-\rho^2)] \right]^{\frac{1}{1-q}}.
\end{equation}
Equation (24) may be rewritten as
\begin{equation}
p_k = \frac{1}{Z_q} {\rm exp}_q\left[ - \zeta
\left(k-\eta \right)^2 \right],
\end{equation}
with
\begin{eqnarray}
Z_q &=& \sum_k \: {\rm exp}_q\left[ - \zeta
\left(k-\eta \right)^2 \right], \\
\zeta &=&\left( \frac{\gamma}{c_q} \right) 
\left[1+\left( \frac{1-q}{c_q} \right)
\left(\frac{\beta^2}{4 \gamma}
+ \beta \mu + \gamma \rho^2 \right) \right]^{-1},\\
\eta &=& - \frac{\beta}{2 \gamma},
\end{eqnarray}
where ${\rm exp}_q(x)$
expresses the $q$-exponential function
defined by
\begin{eqnarray}
{\rm exp}_q(x)&=& [1+(1-q)x]^{\frac{1}{1-q}},
\hspace{2cm}\mbox{for $1+(1-q)x > 0$} \\
&=& 0.
\hspace{5cm}\mbox{for $1+(1-q)x < 0$}
\end{eqnarray}
Parameters of $\zeta$ and $\eta$, newly introduced in place of 
$\beta$ and $\gamma$, are determined 
by the constraints given by Eqs. (20) and (21) as
\begin{eqnarray}
\mu &=& \sum_k P(k)\:k, \\
\rho^2 &=& \sum_k P(k) \:k^2,
\end{eqnarray}
with 
\begin{equation}
P(k) = \frac{ ( {\rm exp}_{q}[-\zeta(k-\eta)^2] )^{q}}
{\sum_k \;({\rm exp}_{q}\:[-\zeta(k-\eta)^2])^{q}}, 
\end{equation}
which are numerically evaluated in general.

\subsection{$Q$-Gaussian}

In order to get some analytical results,
we hereafter assume that $k$ is a continuous variable
varying from $-\infty$ to $\infty$,
which may be justified for $\mu \gg \sigma$.
Equations (31)-(33) yield
\begin{eqnarray}
\mu &=& \eta, \\
\sigma^2 &\equiv& \rho^2 - \mu^2, \\
&=& \left[ \frac{1}{(1-q)\:\zeta} \right] 
\frac{B(\frac{3}{2}, \frac{q}{1-q}+1)}
{B(\frac{1}{2}, \frac{q}{1-q}+1)}, 
\hspace{1.0cm}\mbox{for $q<1$}
\\
&=& \frac{1}{2 \zeta},
\hspace{5.0cm}\mbox{for $q=1$} \\
&=& \left[ \frac{1}{(q-1)\:\zeta} \right] 
\frac{B(\frac{3}{2}, \frac{q}{q-1}-\frac{3}{2})}
{B(\frac{1}{2}, \frac{q}{q-1}-\frac{1}{2})},
\hspace{1.0cm}\mbox{for $q>1$}
\end{eqnarray}
where $B(x,y)$ denotes the beta function.
Equations (35)-(38) lead to
\begin{equation}
\zeta = \frac{1}{2 \nu \sigma^2},
\end{equation}
with
\begin{equation}
\nu = \frac{3-q}{2}. 
\end{equation}
From Eqs. (25), (34) and (39), we get
\begin{equation}
p_k = p_k^Q \equiv \frac{1}{\sqrt{2 \pi}\: \sigma \:D(q,1)} 
\;{\rm exp}_q 
\left[-\frac{ (k-\mu)^2}{2 \nu \sigma^2} \right],
\end{equation}
with
\begin{eqnarray}
D(q,r) 
&=& \sqrt{\frac{\nu}{1-q}}
\;\left( \frac{2 r}{2 r +1 -q} \right)
\; \left( \frac{ \Gamma(\frac{r}{1-q}) }
{ \Gamma(\frac{r}{1-q}+\frac{1}{2}) } \right),
\hspace{1.0cm}\mbox{for $q < 1$} 
\\
&=& 1,
\hspace{8.0cm}\mbox{for $q = 1$} \\ 
&=& \sqrt{\frac{\nu}{q-1}}
\; \left( \frac{ \Gamma(\frac{r}{q-1}-\frac{1}{2}) }
{ \Gamma(\frac{r}{q-1}) } \right),
\hspace{4.0cm}\mbox{for $q > 1$} 
\end{eqnarray}
$\Gamma(x)$ being the gamma function \cite{Note4}.
From Eqs. (22), (23) and (41),
we get degree distribution given by
\begin{equation}
P(k)=P_{Q}(k) \equiv \frac{1}{\sqrt{2 \pi} \:\sigma \: D(q,q) }
\left( {\rm exp}_q\left[-\frac{ (k-\mu)^2}
{2 \nu \sigma^2} \right] \right)^q.
\end{equation}
$p_k^Q$ in Eq. (41) expresses the generalized Gaussian 
which is called $Q$-{\it Gaussian} in this paper.
It is easy to see that 
$D(q,r)$ reduces to unity in the limit of $q=r=1$ because
${\rm lim}_{\mid z \mid \rightarrow \infty} 
\Gamma(z+a)/\Gamma(z) z^a = 1$ \cite{Note3}, then
both $p_k^Q$ and $P_{Q}(k)$ reduce to Gaussian:
\begin{equation}
P_G(k) = P_G(k; \mu, \sigma)
\equiv \frac{1}{\sqrt{2 \pi} \sigma} 
\;{\rm e}^{-\frac{(k-\mu)^2}{2 \sigma^2} },
\end{equation}
with the average $\mu$ and variance $\sigma^2$.
Equations (39) and (40) show that $q \leq 3$ 
because $\nu \geq 0$ for the positive definiteness of 
fluctuations.
We note from Eqs. (29) and (30) that
\begin{eqnarray}
P_{Q}(k) &\propto& (k-\mu)^{-\frac{2q}{q-1}}, 
\hspace{1cm}\mbox{for $\mid k-\mu \mid/\sigma \gg r_c, 
\;q > 1$} \nonumber \\
&=& 0,
\hspace{3cm}\mbox{for $\mid k-\mu \mid/\sigma > r_c, 
\;q < 1$}
\end{eqnarray}
with 
\begin{equation}
r_c = \sqrt{\frac{3-q}{\mid q-1 \mid }}.
\end{equation}

It is worthwhile to note the difference 
between $Q$-Gaussian and the conventional $q$-Gaussian 
which is given by
$p(u) \propto  {\rm exp}_q[- \beta' u^2]$
derived from Eq. (3) with a replacement
of $k \rightarrow (u^2/2)$ \cite{Taruya03}\cite{Lima05},
$u$ standing for the generalized velocity
and $\beta'$ the relevant Lagrange multiplier.
Alternatively, $q$-Gaussian is obtained from $Q$-Gaussian
with $\mu=0$ in Eq. (41).

\subsection{Model calculations}

In order to first get a broad insight to the $Q$-Gaussian,
we show in Figs. 2(a) and 2(b),
some numerical results of $p_k^Q$ and $P_{Q}(k)$
calculated by using Eqs. (41)-(45).
Figure 2(a) shows $p_k^Q$ with $\mu=10$ and $\sigma=1$
for various $q$ values of $q=0.5$, 1.0, 1.5, 2.0 and 2.5.
The $Q$-Gaussian $p_k^Q$ is fat-tailed for $q > 1$ 
and compact support for $q<1$.
Fig. 2(b) shows $P_{Q}(k)$ 
with $\mu=10$ and $\sigma=1$
for various $q$ values corresponding to Fig. 2(a).
We note that a profile of $P_Q(k)$ is rather different
from that of $p_k^Q$: both are normalized by
$\int dk \: p_k^Q= \int dk \:P_Q(k) =1$.
For $q=1.0$, $p_k^Q$ and $P_Q(k)$ reduce to 
the Gaussian distribution located at $k=\mu=10$.
With more increasing $q$ above unity, 
$P_Q(k)$ has a sharper peak
with narrower width. 
In contrast, when $q$ is more decreased below unity,
$P_Q(k)$ has a wider width but has no magnitudes at 
$\mid k-\mu \mid \geq r_c \:\sigma$,
for example, $\mid k-\mu \mid \geq 2.236$ for $q=0.5$. 

\section{Stochastic differential equation approach}

\subsection{Formulation}

Recently Anteneodo and Tsallis \cite{Anten02} have studied
the Langevin equation subject to additive and multiplicative
noises, which lead to nonextensive distributions.
We here assume that Langevin equations are given by
\begin{equation}
\frac{d k_i}{dt}=-\lambda\: (k_i-\mu)+ g(k_i)\;\xi_i(t)
+ \eta_i(t),
\hspace{1.0cm}\mbox{($i=1-N$)}
\end{equation}
where $k_i(t)$ denotes a real variable
expressing the number of couplings of the node $i$:
$\lambda$ the relaxation rate:
$g(k)$ an arbitrary function whose explicit form 
will be given later [Eqs. (52), (88) and (89)]:
$\eta_i(t)$ and $\xi_i(t)$ are additive and
multiplicative white noises,
respectively, as given by
\begin{eqnarray}
<\eta_i(t) \:\eta_j(t')> &=& 2 A \:\delta_{ij}\:\delta(t-t'),\\
<\xi_i(t) \:\xi_j(t')> &=& 2 M \:\delta_{ij}\:\delta(t-t'),
\end{eqnarray}
means of noises being vanishing.
Equation (49) shows that the number of couplings $k_i$
is influenced by both additive and multiplicative noises.
In particular, effects of multiplicative noises given by $g(k_i)$
depend on the present state of node $k_i(t)$.

When $g(k)$ is given by
\begin{equation}
g(k)=k-\mu, 
\end{equation}
we get the Fokker-Planck equation 
for $p(k_i,t)\;(\equiv p(k,t))$ given by \cite{Anten02}
\begin{equation}
\frac{\partial p(k,t)}{\partial t} 
= \lambda \;\frac{\partial}{\partial k} \left[g(k) \:p(k,t) \right]
+ M \;\frac{\partial}{\partial k}
\left( g(k)\:\frac{\partial}{\partial k} [g(k) \:p(k,t)] \right)
+ A \;\frac{\partial^2 p(k,t)}{\partial k^2}.
\end{equation}
Its stationary solution is given by
\cite{Anten02}\cite{Sakaguchi01}
\begin{equation}
p_k = p(k,\infty) \propto \left[1+\left( \frac{M}{A} \right)
(k-\mu)^2
\right]^{-\frac{\lambda+M}{2M}}.
\end{equation}
Equation (54) may be rewritten 
in two ways as
\begin{eqnarray}
p_k &\propto& \left[1-(1-q)
\frac{(k-\mu)^2}{2 \overline{\sigma}^2}
\right]^{\frac{1}{1-q}}, \\
&\propto& \left[1-(1-q) 
\frac{(k-\mu)^2}{2 \nu \sigma^2}
\right]^{\frac{1}{1-q}},
\end{eqnarray}
with 
\begin{eqnarray}
q &=& \frac{\lambda + 3M}{\lambda+M}, \\
\overline{\sigma}^2 &=& \frac{A}{\lambda+M}, \\
\sigma^2 &=& \frac{A}{\lambda},
\end{eqnarray}
where $\nu$ is given by Eq. (40).
Note the difference between $\overline{\sigma}$ in Eq. (58)
and $\sigma$ in Eq. (59).
The $Q$-Gaussian given by Eq. (55) with $\mu=0$ is adopted in
\cite{Anten02}. Equation (56)
agrees with the result obtained by MEM
given by Eq. (41) for $q \geq 1$.

Equation (57) shows that
the nonextensivity arises from
multiplicative noises.
In the case of no multiplicative noises ($M=0$),
we get $q=1$, leading to the Gaussian distribution:
\begin{equation}
p_k =P_G(k)
=\frac{1}{\sqrt{2 \pi} \sigma} 
\;{\rm e}^{-\frac{(k-\mu)^2}{2 \sigma^2} }.
\end{equation}

\subsection{Model calculations}

We show model calculations, simulating
Langevin equation given by Eq. (49) 
with $N=100$: the Heun method is employed with
a time step of 0.0001.
Figures 3(a) and 3(b) show time courses of
$k_i(t)$ [$\equiv k(t)$] for $(A, M)=(1.0, 0.0)$ and 
(1.0, 0.5), respectively, with $g(k)=k-\mu$,
$\lambda=1.0$ and $\mu=10$.
The number of $k$, starting from
an initial value of $\mu=10$, fluctuates around $k=\mu$
by effects of noises. 
A comparison between Figs. 3(a) and 3(b) shows
that fluctuations are much increased by effects
of multiplicative noises.
Circles in Figs. 4(a) and 4(b) show
degree distributions $P(k)$ 
for $(A, M)=(1.0, 0.0)$ and
(1.0, 0.5), respectively, obtained 
by averages over $N=100$ for $50 < t \leq 100$ \cite{Note5}. 
Solid curves in Figs. 4(a) and 4(b) express
$P_Q(k)$ given by Eq. (45) for $(q, \sigma)=(1.0, 1.008)$
and (1.667, 0.984), respectively:
adopted $q$ values are calculated by Eq. (57) and $\sigma$
are evaluated by using $P_{Q}(k)$ obtained by simulations
\cite{Note5}.
Degree distribution for $M=0$ in Figs. 4(a) shows 
Gaussian. When the multiplicative noises
with $M=0.5$ are introduced, degree distribution 
$P(k)$ has tails expressed by the power law as shown in Fig. 4(b).
Results of stochastic differential equation are in good agreement
with those of $Q$-Gaussian shown by solid curves.

\section{Hidden-parameter distribution approach}

Employing Poisson distribution $P_P(k)$
for random networks, 
Abe and Thurner \cite{Abe05} proposed
the hidden-parameter expansion to get 
\begin{equation}
\int_0^{\infty} \: d \mu' \; P_P(k;\mu')\:\Pi_1(\mu')  
= P(k),
\end{equation}
where $\Pi_1(\mu)$ stands for distribution of
the hidden variable $\mu$.
For scale-free distribution $P_{SF}(k)$ given by
\begin{eqnarray}
P_{SF}(k) &=& A \:(k+k_0)^{-\gamma}, 
\end{eqnarray}
they obtained \cite{Abe05} 
\begin{eqnarray}
\Pi_1(\mu) &\propto& \mu^{-\delta}, 
\hspace{1cm}\mbox{for large $\mu$}
\end{eqnarray}
with 
\begin{equation}
\delta=\gamma,
\end{equation}
where $\gamma > 1$, and
$A$ and $k_0$ denote constants.

It should be noted that degree distribution of 
random networks may be described not only by Poisson $P_P(k)$
but also by Gaussian distribution $P_G(k)$.
$P_P(k)$ is obtained for large $N$ from binomial distribution 
$P_B(k)$  as shown by Eq. (5).
Alternatively, $P_G(k)$ may be derived from binomial distribution
when $k$ in $P_B(k)$ is assumed to be a real variable as follows.
Expanding $P_B(k)$ around $k = \mu = N c$
where $P_{B}(k)$ takes the maximum value, 
we get
\begin{eqnarray}
P_B(k) &=& {\rm e}^{{\rm ln} P_B(k)}
\simeq {\rm e}^{{\rm ln} P_B(\mu)-\frac{(k-\mu)^2}{2 \sigma^2}}, 
\nonumber \\
&=& \frac{1}{\sqrt{2 \pi} \sigma}
\; {\rm e}^{{-\frac{(k-\mu)^2}{2 \sigma^2}}}=P_G(k), 
\end{eqnarray}
where $\sigma^2=Nc(1-c)$.

When generalizing
the hidden-parameter method proposed 
by Abe and Thurner \cite{Abe05},
we may assume that $P(k)$ is
expressed as a superposition of Gaussians $P_G(k)$
for random networks, as 
\begin{equation}
\int_0^{\infty} \: d \mu' \;
\int_0^{\infty} \: d \sigma'\; P_G(k;\mu',\sigma')
\:\Pi(\mu', \sigma') 
= P(k),
\end{equation}
where $\Pi(\mu, \sigma)$ denotes distribution for
hidden variables $\mu$ and $\sigma$.
When $P_{SW}(k)$ obtained by simulations or
an analytical expression given by Eq. (6)
is substituted to $P(k)$ in Eq. (66),
$\Pi(\mu,\sigma)$ will be determined 
by inversely solving Eq. (66). 
This would be interesting 
but seems rather difficult.

When we assume 
\begin{equation}
\Pi(\mu',\sigma')=\Pi_2(\sigma')\:\delta(\mu'-\mu),
\end{equation}
Eq. (66) becomes
\begin{equation}
\int_0^{\infty} \:d \sigma'
\:P_G(k;\mu,\sigma')
\:\Pi_2(\sigma')  
= P(k),
\end{equation}
where $\mu=<k>$.
We may get $\Pi_2(\sigma')$ for $P_{SW}(k)$ given by Eq. (6),
in principle, from Eq. (68), which
is, however, problematic because
$P_{SW}(k)$ is valid only for $k \geq \mu/2$.

Instead, we have obtained $\Pi_2(\sigma)$
for $Q$-Gaussian $p_k^Q$ from Eq. (68) by
\begin{equation}
\int_0^{\infty} \: d \sigma'
\:\frac{1}{\sqrt{2 \pi} \sigma'} 
\;{\rm e}^{-\frac{(k-\mu)^2}{2 \sigma'^2}} \;\Pi_2(\sigma') 
= {\rm exp}_q\left[ -\frac{(k-\mu)^2}{2 \nu \sigma^2} \right],
\end{equation}
where $\nu=(3-q)/2$ [Eq. (40)],
and normalization factor in $p_k^Q$
is neglected for a simplicity of calculation.
By using the change of variables:
$1/2 \sigma'^2=\alpha$,  
$1/2 \nu \sigma^2=\alpha_0$,
and $(k-\mu)^2=x$,
we get
\begin{equation}
\int_0^{\infty} \: d \alpha  
\;{\rm e}^{-\alpha x}\:f(\alpha) 
= {\rm exp}_q(-\alpha_0 x),
\end{equation}
with
\begin{eqnarray}
f(\alpha)&=& \frac{1}{\sqrt{2 \pi}}
\:\left( \frac{1}{2 \alpha} \right)
\Pi_2\left( \left(\frac{1}{2 \alpha} \right)^{1/2} \right),\\
&=& \frac{1}{\Gamma \left( \frac{n}{2} \right)}
\left( \frac{n}{2\alpha_0} \right)^{\frac{n}{2}}
\alpha^{\frac{n}{2}-1} 
\;e^{ -\frac{n \alpha}{2\alpha_0} }, \\
\alpha_0&=& \frac{1}{2 \nu \sigma^2} =E(\alpha), \\
q-1&=&\frac{2}{n}= \frac{E(\alpha^2)-E(\alpha)^2}{E(\alpha)^2},
\end{eqnarray}
where $f(\alpha)$ in Eq. (72) denotes the $\Gamma$- 
(or $\chi^2$-) distribution function
of the order $n$, $E(\alpha)$ the average 
of the variable $\alpha$ over $f(\alpha)$, and
$E(\alpha^2)$ its variance
\cite{Wilk00}\cite{Beck02}.
Equation (70) is nothing but the Laplace transformation.
By using the relation derived from Eq. (71):
\begin{eqnarray}
\Pi_2(\sigma) &=& \sqrt{2 \pi}\:\left( \frac{1}{\sigma^2} \right)
\;f\left( \frac{1}{2\sigma^2} \right),
\end{eqnarray}
we get 
\begin{eqnarray}
\Pi_2(\sigma) 
&\propto& \sigma^{- \delta},
\hspace{1cm}\mbox{for large $\sigma$} 
\end{eqnarray}
with
\begin{equation}
\delta=\gamma  = \frac{2}{q-1}, 
\end{equation}
where $\gamma$ denotes the index for the expression given by
\begin{eqnarray}
p_k^Q &\propto& (k-\mu)^{-\gamma}.
\hspace{1cm}\mbox{for large $k$}
\end{eqnarray}
Eqs. (76)-(78) should be compared
to the result of Ref. \cite{Abe05}
given by Eqs. (62)-(64),
related discussion being given in Sec. 6.

Figure 5 shows Log plots of $\Pi_2(\sigma)$ for
$Q$-Gaussian with various $q$ values
against $\sigma/(\sigma_0 \sqrt{\nu})$ 
where $\sigma_0=1/\sqrt{2 \alpha_0}$ for $q=1.0$.
In the limit of $q \rightarrow 1.0$ ($n \rightarrow \infty$),
$\Pi_2(\sigma)$ becomes the delta function 
at $\sigma=\sigma_0$. 
As $q$ is more increased than unity,
$\Pi_2(\sigma)$ has a wider width, expressing larger 
fluctuations in the variable $\sigma$.
For $q=1.5$, 2.0 and 2.5, the index $\delta$ given by Eq. (77)
is 4, 2 and 1.33, 
whose slopes are shown by chain curves in Fig. 5.

If we express Gaussian in Eq. (66) as
\begin{equation}
P_G(k;\mu',\sigma') 
= \psi\left( \frac{k-\mu'}{\sigma'} \right),
\end{equation}
we get
\begin{eqnarray}
P(k)&=& \int_0^{\infty} \: d \mu' 
\;\int_0^{\infty} \: d \sigma'
\;\psi\left( \frac{k-\mu'}{\sigma'} \right)
\:\Pi(\mu', \sigma'),
\end{eqnarray}
which is similar to the wavelet transformation \cite{Wavelet}.
It is interesting
to obtain analytical
expression of $\Pi(\mu, \sigma)$ 
for a general $P(k)$.


\section{Analysis of degree distribution in SW networks}

We will discuss an application of the $Q$-Gaussian 
to an analysis of $P_{SW}(k)$.
The coupling connectivity $R$ has been calculated
with the use of Eq. (10) by simulations (1000 trials)
for SW networks 
with $N=100$ and $\mu=10$ as a function of $p$
\cite{Hasegawa04}.
We have obtained $\sigma$
by using $\sigma =\mu \sqrt{R-1}$ derived from
Eqs. (21) and (35).
Chain and dashed curves in Fig. 6 show 
the $p$ dependence of
$R$ and $\sigma$, respectively. 
With increasing $p$, both $R$ and $\sigma$
are monotonously increased.

We have tried to fit $P_{Q}(k)$ in Eqs. (41)-(45) to
$P_{SW}(k)$ obtained by simulations,   
choosing a proper $q$ value  
with $\mu$ and $\sigma$ calculated for a given $p$ shown in Fig. 6.
When we have chosen the $q$ value 
such that maxima of the two distributions agree,
their whole shapes are in good agreement.
Figures 7(a)-7(f) show such fittings for $p=0.1$,
0.2, 0.4, 0.6, 0.8 and 1.0.
In the case of $p=0.1$ shown in Fig. 7(a), 
$P_{Q}(k)$ with $q=1.35$, $\sigma=0.97$ and $\mu=10$
well reproduces the relevant degree distribution
plotted by circles.
For a comparison, we show also results with
the use of $q=1.5$ and 1.2
whose agreements are worse than that of $q=1.35$.
Gaussian ($q=1.0$) for $\mu=10$ and $\sigma=0.97$ is 
in significant disagreement with $P_{SW}(k)$
at $9 \leq k \leq 11$.
In the case of $p=0.2$ shown in Fig. 7(b),
$P_Q(k)$ with $q=1.2$, $\sigma=1.34$ and $\mu=10$
well explains $P_{SW}(k)$ obtained by simulations,
results with $q=1.0$ and 1.4 being also plotted
for a comparison.
In cases of $p=0.4$, 0.6 and 0.8 shown in 
Figs. 7(c), 7(d) and 7(e),
we have obtained a good agreement between $P_{SW}(k)$ and $P_Q(k)$ 
with $q=1.0$ expressing the Gaussian distribution.
In contrast, in the case of $p=1.0$ shown 
in Fig. 7(f),
the result of $q=0.95$ yields a better fit
than that of $q=1.0$.

At a glance at Figs. 7(a)-7(f), we have an impression that
$P_{SW}(k)$ is well described by $P_{Q}(k)$.
It is, however, not true in a strict sense,
because at $\mid k-\mu \mid/\sigma \gg r_c$,
$P_{SW}(k)$ follows 
the exponential function
while $P_{Q}(k)$ obeys the power law ($q > 1$),
as Eq. (47) shows.
This is clearly seen in Fig. 8(a)-8(f) where
results in Figs. 7(a)-7(f) are shown in the Log plots.
For example,
in the case of $p=0.1$,
the agreement of $P_{Q}(k)$ with $P_{SW}(k)$ is good for 
$\mid k-\mu \mid \siml 3$ as demonstrated 
in Fig. 7(a), but it becomes worse for 
$\mid k-\mu \mid  \simg 3 $.
Nevertheless, it should be stressed that {\it main parts}
of $P_{SW}(k)$ are well described by {\it a single} $P_{Q}(k)$
as shown in Fig. 7(a)-7(f). 
When $P_{SW}(k)$ and $P_{Q}(k)$ are in good agreement
at $k \in [\mu-k_a,\:\mu+k_a]$
with an accuracy $\mid P_{SW}(k)-P_Q(k) \mid \leq 10^{-2}$, 
the number of $k$ states
included in this region is given by
\begin{equation}
n_a = \int_{\mu-k_a}^{\mu+k_a} dk \;P(k),
\end{equation}
which yields $n_a=0.992$
for $k_a = 3$ in the case of $p=0.1$. 
Similarly, we get 
$(p, k_a, n_a)=(0.2, 4, 0.993)$, (0.4, 4, 0.975), 
(0.6, 5, 0.983), (0.8, 5, 0.970), and (1.0, 5, 0.959).

\section{Discussions}

We will present more detailed analysis, 
trying to reduce the discrepancy between
$P_{SW}(k)$ and $P_Q(k)$ at their tails in this section.
$P_{SW}(k)$ calculated by simulations for SW networks
with $p=0.1$, $\mu=10$ and $N=100$
is again shown by filled circles in Fig. 9(a), 
where the analytical result valid for large $N$ given by Eq. (6)
is also shown by filled triangles for a comparison.
We note a deviation of $P_{SW}(k)$ from $P_Q(k)$ 
shown by solid curve at $\mid k-\mu \mid \simg 3$,
where $P_{SW}(k)$ obeys the exponential (Gaussian) law. 
This suggests that $p_k$ in SW networks behaves as  
\begin{eqnarray}
p_k 
&\propto& {\rm exp}_{q}
\left[ -\frac{(k-\mu)^2}{(3-q) \sigma^2} \right],
\hspace{1cm}\mbox{for small $\mid k-\mu \mid$ } \nonumber \\
&\propto& {\rm e}^{-\frac{(k-\mu)^2}{2 \sigma'^2}}.
\hspace{3cm}\mbox{for large $\mid k-\mu \mid$ } 
\end{eqnarray}

If we incorporate the $k$-dependence 
into the entropic index,
the entropy in Eq. (18) is generalized as
\begin{equation}
S_{q}=\frac{\sum_k p_k^{q_k} - 1}{1-q_k},
\end{equation} 
where $q_k$ denotes the $k$-dependent entropic index.
MEM with the constraints
given by Eqs. (19)-(21) with a replacement of 
$q \rightarrow q_k$ yields
\begin{equation}
p_k = \frac{1}{Z_q}  
\;{\rm exp}_{q_k}
\left( \frac{(k-\mu)^2}{(3-q_k)\sigma^2} \right),
\end{equation}
with
\begin{equation}
Z_q = \int d k  
\; {\rm exp}_{q_k}
\left( \frac{(k-\mu)^2}{(3-q_k)\sigma^2} \right),
\end{equation}
from which $P(k)$ is given by Eqs. (22) and (23) with 
a replacement of $q \rightarrow q_k$.
We expect that the result given by Eq. (82) may
be realized in Eq. (84) if $q_k=q$ for small $\mid k-\mu \mid$
and $q_k=1$ for large $\mid k-\mu \mid$.

Bearing these consideration in mind, we have assumed
the $k$-dependent entropic index given by
\begin{eqnarray}
q_k &=& 1 + (q-1) \:\Theta(k-\mu+k_1,\theta)
\; \Theta(\mu+k_1-k,\theta), 
\end{eqnarray}
with 
\begin{eqnarray}
\Theta(k,\theta) &=& \frac{1}{ {\rm e}^{(- k/\theta)} +1},
\end{eqnarray}
where $k_1$ denotes the characteristic $k$ value.
Equation (86) shows that
$q_k=q$ for $\mid k-\mu \mid \siml k_1$ and
$q_k=1$ for $\mid k-\mu \mid \simg k_1$.
Open squares in Fig. 9(a) show $P(k)$ 
calculated by using $q_k$ given by Eqs. (86) and (87)
with $k_1=4$ and $\theta=1$.
The $k$-dependence of $P(k)$
changes from $Q$-Gaussian to Gaussian with increasing
$\mid k-\mu \mid$, and it is in better agreement
with $P_{SW}(k)$ than $P_Q(k)$.

In order that
the $k$-dependence in the entropic index given by Eq. (86)
is realized in the Langevin model,
we have tried to modify a function $g(k)$ for multiplicative noises.
Open circles in Fig. 9(b) express the results of 
simulations of Langevin equation (49) 
with $g(k)=k-\mu$ given by Eq. (52) 
for $\lambda=1$, $A=0.941$ and $M=0.212$,
which are chosen from Eqs. (57) and (59) 
with $q=1.35$ and $\sigma=0.97$
for $p=0.1$ \cite{Note5}. 
In contrast, 
open squares in Fig. 9(b) denote the result of
simulations of the Langevin model
with $g(k)$ given by
\begin{eqnarray}
g(k) &=& (k-\mu) \:\Theta(k-\mu+k_2,\theta)
\; \Theta(\mu+k_2-k,\theta), 
\end{eqnarray}
for $\lambda=1$, $\mu=10$, $A=0.941$, $M=0.212$,  
$k_2=4$ and $\theta=1$. 
The result shown by
open squares in Fig. 9(b) 
deviates from
that shown by open circles 
at $\mid k-\mu \mid \simg 4$ where
effects of multiplicative noises do not work
for $g(k)$ given by Eq. (88).
It is shown that 
an agreement of $P_{SW}(k)$ with
$P(k)$ expressed by open squares 
is better 
than that with $P_Q(k)$ expressed by the solid curve 
(and open circles).

A similar analysis has been made for the case of $p=0.2$,
whose results are shown in Figs. 10(a) and 10(b).
Filled circles and triangles in Fig. 10(a) denote $P_{SW}(k)$
obtained by simulations and analytic method, respectively.
Solid curve, expressing $P_Q(k)$ with $q=1.20$
and $\sigma=1.34$, deviates from $P_{SW}(k)$
at $\mid k-\mu \mid \simg 4$. In contrast,
open squares express the result calculated with the use of $q_k$
given by Eq. (86) for $k_1=5$ and $\theta=1$,
which are in better agreement
with $P_{SW}(k)$ than $P_Q(k)$.

SDE has been applied to the case of $p=0.2$.
Open circles in Fig. 10(b) express the result
of simulations for Langevin equations with $g(k)=k-\mu$, 
$\lambda=1$, 
$\mu=10$,
$A=1.788$ and $M=0.111$, which are chosen 
from Eqs. (57) and (59)
with $q=1.20$ and $\sigma=1.34$ for $p=0.2$ \cite{Note5}.
In contrast, open squares in Fig. 10(b) denote $P(k)$ 
calculated by simulation with
$g(k)$ given by Eq. (88) for 
$\lambda=1$, $\mu=10$,
$A=1.788$, $M=0.111$,
$k_2=5$ and $\theta=1$.
An agreement of $P_{SW}(k)$ with $P(k)$ is much better than
that with $P_Q(k)$ shown by the solid curve 
(and open circles).

We have obtained the optimum entropic index of $q=1.0$
for $p=0.4$, 0.6 and 0.8 [Fig. 7(c)-7(e)],
and $q=0.95$ for $p=1.0$ [Fig. 7(f)].
Log plots in Figs. 8(c)-8(f) show that
there are deviations between $P_{SW}(k)$ and $P_Q(k)$
at $\mid k-\mu \mid \simg 5$. 
A Log plot of results of $p=1.0$
are again shown in Fig. 11, where $P_{SW}(k)$
calculated by simulations and analytical method
are shown by filled circles and triangles, respectively.
We note that at $\mid k-\mu \mid \simg 4$,
$P_{SW}(k)$ becomes gradually larger than $P_Q(k)$
for $q=0.95$ and $\sigma=6.0$
shown by solid curve.
In order to reduce the discrepancy at $\mid k-\mu \mid \simg 4$,
we adopt a function of $g(k)$ for multiplicative noises
given by
\begin{eqnarray}
g(k) &=& \Theta(k-\mu-k_3, \theta),
\end{eqnarray}
which has a magnitude of unity at $k \simg \mu+k_3$
and zero at $k \siml \mu+k_3$.
This expression with Eq. (49) may be 
alternatively interpreted such that 
magnitude of {\it additive} noises is increased 
by an amount of $M$ at $k \simg \mu+k_3$.
Open circles in Fig. 11 show $P(k)$ calculated 
by using SDE with $g(k)=k-\mu$, $\lambda=1$,
$A=6.0$ and $M=0$, which are chosen from Eqs. (57) and (59)
with $q=1.0$ and $\sigma=2.45$ \cite{Note5}.
In contrast, open squares express $P(k)$ 
calculated with $g(k)$ given by Eq. (89)
with $\lambda=1$, $A=6.0$, $M=2.0$, $k_3=4$ and $\theta=1$, which
is in better agreement with $P_{SW}(k)$
than $P_Q(k)$ shown by the solid curve.

It is necessary to point out that
the normalization
factor of $1/\sqrt{2 \pi} \sigma'$ in Eq. (69) plays
an important role in determining $\Pi_2(\sigma')$
as shown below.
When $Q$-Gaussian of the right-hand side in Eq. (69) is replaced by
\begin{equation} 
P(k)=(k-\mu)^{-\gamma},
\hspace{1cm}\mbox{($\gamma > 1$)}, 
\end{equation}
we get
\begin{equation}
\int_0^{\infty} \: d \alpha  
\;{\rm e}^{-\alpha x}\:f(\alpha) 
= x^{-\gamma/2},
\end{equation}
with $1/2 \sigma'^2=\alpha$, $(k-\mu)^2=x$, and
\begin{eqnarray}
f(\alpha)
&=& \frac{1}{\sqrt{2 \pi}}
\:\left( \frac{1}{2 \alpha} \right)
\Pi_2\left( \left(\frac{1}{2 \alpha} \right)^{1/2} \right),\\
&=& \frac{\alpha^{\gamma/2-1}}{\Gamma(\gamma/2)},
\end{eqnarray}
which lead to
\begin{equation}
\Pi_2(\sigma) \propto \sigma^{- \gamma},
\hspace{1cm}\mbox{for large $\sigma$}
\end{equation}
as given by Eqs. (76) and (77).

On the contrary, if the normalization factor of 
$1/\sqrt{2 \pi} \sigma'$
is neglected in Eq. (76), we get Eqs. (91) and (93) with
\begin{eqnarray}
f(\alpha)&=& \left( \frac{1}{2 \alpha} \right)^{3/2}
\Pi_2\left( \left(\frac{1}{2 \alpha} \right)^{1/2} \right).
\end{eqnarray}
By using the relation obtained from Eq. (95): 
\begin{eqnarray}
\Pi_2(\sigma) &=& 
\:\left( \frac{1}{\sigma^3} \right)
\;f\left( \frac{1}{2\sigma^2} \right),
\end{eqnarray}
we get
\begin{eqnarray}
\Pi_2(\sigma) &\propto& \sigma^{-(\gamma+1)}.
\hspace{1cm}\mbox{for large $\sigma$}
\end{eqnarray}
The difference between indices of $\sigma$ 
in Eqs. (94) and (97) arises from 
the normalization factor $1/\sqrt{2 \pi} \sigma'$
of Gaussian in Eq. (69).

\section{Conclusions}

In summary, degree distribution 
of WS small-world networks has been discussed 
in terms of $Q$-Gaussian obtained by the
three approaches: MEM, SDE, and HVD. 
It is interesting that 
with an accuracy of
$\mid P_{SW}(k)-P_Q(k)\mid \siml 10^{-2} $,
a {\it single} $P_Q(k)$ 
may describe main parts of $P_{SW}(k)$ 
which is expressed as a superposition of Poissons in Eq. (6).
In order to account for
the overall behavior in $P_{SW}(k)$ including 
its main part and tails, we have assumed 
the $k$-dependent entropic index in MEM with
the entropy given by Eq. (83), which is nonextensive
unless $q_k=1$ for all $k$.

This $k$-dependent entropic index is realized in SDE, by adopting
multiplicative noises
given by Eq. (88) for $p=0.1$ and 0.2, and those
given by Eq. (89) for $0.4 \siml p \siml 1.0$.
Degree distributions 
of scale-free and random networks are obtained
for $g(k)=k-\mu$ and $g(k)=0$, respectively, in SDE.
We expect 
that $P(k)$ of SW networks has the combined property
of scale-free and random networks because
$g(k)$ is $k-\mu$ and 0 for small $\mid k-\mu \mid $ and 
large $\mid k-\mu \mid $,
respectively.
It is necessary to understand the relation
between rewiring processes in making WS-SW networks \cite{Wat98}
and the two types of multiplicative noises in Langevin
equation. 

Although the three approaches, MEM, SDE, and HVD,
are conceptually rather different,
they lead to equivalent results.
There are, however, some differences among them.
(i) The entropic index $q$ is given 
by $q=1+2M/(\lambda+M)$ and $q=1+2/n$
in SDE and HVD, respectively, while that in MEM 
is a free parameter with the constraint of $q \leq 3$. 
(ii) The escort probability $P(k)$ is obtained
from $p_k$ as given by Eqs. (22) and (23) in MEM,
while there are no ways in SDE and HVD.
Although no analyses have been made for $P_{SW}(k)$ 
with the use of HVD, Gaussian-based expression 
(66) or (81) is expected to be promising.
This subject is left as our future study.

\section*{Acknowledgements}
This work is partly supported by
a Grant-in-Aid for Scientific Research from the Japanese 
Ministry of Education, Culture, Sports, Science and Technology.  



\newpage

\begin{figure}
\caption{
Degree distribution $P_{SW}(k)$ (filled marks) 
of the WS small-world model
with $N=100$ and $\mu=10$ for various $p$ values,
calculated by simulations with 1000 trials,
solid lines being shown for a guide of the eye.
As a comparison, $P(k)$ for random
networks with the same parameters is plotted
by open circles (see text).
}
\label{fig1}
\end{figure}

\begin{figure}
\caption{
(a) The $Q$-Gaussian $p_k^Q$ and 
(b) $P_{Q}(k)$ with $\mu=10$ and $\sigma=1.0$
for various $q$ values. 
}
\label{fig2}
\end{figure}

\begin{figure}
\caption{
Time courses of $k(t)$ obtained by dynamical simulations
of stochastic differential equations [Eqs. (49)-(52)] 
for (a) $A=1.0$ and $M=0.0$, and 
(b) $A=1.0$ and $M=0.5$ with $\lambda=1.0$ and $\mu=10$.
Results are simulations of a single trial
by using the Heun method
with the time step of 0.0001. 
}
\label{fig3}
\end{figure}

\begin{figure}
\caption{
(color online)
Log-plots of $P(k)$ calculated by simulations (circles)
of stochastic differential equations [Eqs. (49)-(52)] 
for (a) $A=1.0$ and $M=0.0$, and 
(b) $A=1.0$ and $M=0.5$ with $\lambda=1.0$ and $\mu=10$,
and those of $P_{Q}(k)$ [Eq. (45)] (solid curves):
$q=1.0$ and $\sigma=1.008$ for (a) 
and $q=1.667$ and $\sigma=0.984$ for(b). 
Results of simulations are averages of $N=100$
for $50 < t \leq 100$  by using
the Heun method with the time step of 0.0001.
}
\label{fig4}
\end{figure}

\begin{figure}
\caption{
Log plots of $\Pi_2(\sigma)$
for $Q$-Gaussian against $\sigma/(\sigma_0 \:\sqrt{\nu})$
with various $q$ values, 
their slopes for $q=1.5$, 2.0 and 2.5 being shown by 
chain curves with $\delta=4$, 2 and 1.33, respectively
($\sigma_0=1/\sqrt{2 \alpha_0}$ for $q=1.0$).
}
\label{fig5}
\end{figure}

\begin{figure}
\caption{
The $p$ dependence of $R$ (chain curve),
$\sigma$ (dashed curve) and $q$ (solid curve) adopted for 
a fitting of $P_{Q}(k)$ to $P_{SW}(k)$ of simulations
for SW networks with $N=100$ and $\mu=10$,
$R-1$ being multiplied by a factor of ten.
}
\label{fig6}
\end{figure}

\begin{figure}
\caption{
(color online)
$P_{SW}(k)$ with $N=100$ and $\mu=10$
obtained by simulations (circles), 
and $P_{Q}(k)$ (solid curves) with adopted $q$ values
for (a) $p=0.1$, (b) $p=0.2$, (c) $p=0.4$,
(d) $p=0.6$, (e) $p=0.8$, and (f) $p=1.0$.
Dashed lines are shown for a guide of the eye.
}
\label{fig7}
\end{figure}

\begin{figure}
\caption{
(color online)
Log-plots of Fig. 7.
}
\label{fig8}
\end{figure}

\begin{figure}
\caption{
(color online)
(a) Log plots of
$P_{SW}(k)$ for $p=0.1$ with $\mu=10$ and $N=100$ 
calculated by simulations 
(filled circles) and analytical method (filled triangles),
$P_Q(k)$ (solid curve),
and $P(k)$ with $q_k$ given by Eq. (86) for $k_1=4$
and $\theta=1$ (open squares).
(b) $P(k)$ in Langevin method with $g(k)=k-\mu$, $\lambda=1$,
$A=0.941$ and $M=0.212$ (open circles),
and that with $g(k)$ given by Eq. (88), $\lambda=1$,
$A=0.941$, $M=0.212$,
$k_2=4$ and $\theta=1$ (open squares) (see text).
}
\label{fig9}
\end{figure}

\begin{figure}
\caption{
(color online)
(a) Log plots of 
$P_{SW}(k)$ for $p=0.2$ with $\mu=10$ and $N=100$ 
calculated by simulations 
(filled circles) and analytical method (filled triangles), 
$P_Q(k)$ (solid curve),
and $P(k)$ with $q_k$ given by Eq. (86) for $k_1=5$
and $\theta=1$ (open triangles).
(b) $P(k)$ in Langevin method with $g(k)=k-\mu$, $\lambda=1$,
$A=1.788$ and $M=0.111$ (open circles),
and that with $g(k)$ given by Eq. (88), $\lambda=1$,
$A=1.788$, $M=0.111$,
$k_2=5$ and $\theta=1$ (open squares) (see text).
}
\label{fig10}
\end{figure}

\begin{figure}
\caption{
(color online)
Log plots of
$P_{SW}(k)$ for $p=1.0$ with $\mu=10$ and $N=100$ 
obtained by simulations (filled circles) and 
analytical method (filled triangles), 
$P_Q(k)$ with $q=0.95$ and $\sigma=2.45$ (solid curve),
$P(k)$ in Langevin method with $g(k)=k-\mu$, $\lambda=1$,
$A=6.0$ and $M=0$ and (open circles),
and that with $g(k)$ given by Eq. (89), $\lambda=1$,
$A=6.0$, $M=2.0$, 
$k_3=4$ and $\theta=1$ (open squares) (see text).
}
\label{fig11}
\end{figure}

\end{document}